\documentclass[%
 reprint,
 amsmath,amssymb,
 aps,
pre,
]{revtex4-2}

\usepackage{subcaption}
\usepackage[version=4]{mhchem}
\usepackage{graphicx}% Include figure files
\usepackage{dcolumn}% Align table columns on decimal point
\usepackage{bm}% bold math
\usepackage[hidelinks]{hyperref}
\def\equationautorefname#1#2\null{%
  Eq.#1(#2\null)%
}
\def\tableautorefname#1#2\null{%
  {table#1#2\null}%
}
\def\sectionautorefname#1#2\null{%
  {Section#1#2\null}%
}
\def\figureautorefname#1#2\null{%
  {Fig.#1#2\null}%
}
\def\subsectionautorefname#1#2\null{%
  {Section#1#2\null}%
}
\usepackage{float}
\usepackage{braket}
\usepackage{mathtools}

\usepackage{xcolor} %Add additional comments
\newcommand{\add}[1]{\textcolor{black}{#1}}

\begin{document}

\captionsetup{justification=raggedright}

\preprint{APS/Ununterscheidbare Links}

\title{An estimator of entropy production for partially accessible Markov networks based on the observation of blurred transitions}% Force line breaks with \\

\author{Benjamin Ertel}%
\author{Udo Seifert}%
\affiliation{%
 II. Institut für Theoretische Physik, Universität Stuttgart, 70550 Stuttgart, Germany
}%

\date{\today}% It is always \today, today,

\begin{abstract}
A central task in stochastic thermodynamics is the estimation of entropy production for partially accessible Markov networks. We establish an effective transition-based description for such networks with transitions that are not distinguishable and therefore blurred for an external observer. We demonstrate that, in contrast to a description based on fully resolved transitions, this effective description is typically non-Markovian at any point in time. \add{Starting from an information-theoretic bound, we derive an operationally accessible entropy estimator for this observation scenario.} We illustrate the operational relevance and the quality of this entropy estimator \add{with a} numerical analysis of various representative examples.    
\end{abstract}

\maketitle

%\tableofcontents

\section{Introduction}
One major result of stochastic thermodynamics is the identification of entropy production for physical systems within a Markovian description \cite{sekimoto2010,jarzynski2011,seifert2012}. Based on this identification, the dissipation in chemical and biophysical systems, for instance chemical reaction networks \cite{schmiedl2007,ge2012,rao2016}, can be quantified theoretically. In practice, however, many systems are only partially accessible, i.e., the full Markovian description is not \add{observed}, implying that the entropy production is not directly operationally accessible. 

For inferring the entropy production of these partially accessible systems, various strategies with different level of sophistication have been proposed. The apparent entropy production \cite{esposito2012,mehl2012,uhl2018,bo2017} and state-lumping entropy estimators \cite{rahav2007,gomez-marin2008,puglisi2010,bo2014,seiferth2020} bound the full entropy production with the coarse-grained entropy production of effective Markov models. Time-series estimators \cite{roldan2010,roldan2012} and fluctuation theorem estimators \cite{shiraishi2015,polettini2017,bisker2017} provide entropy bounds based on the irreversibilty of time-antisymmetric observables. The thermodynamic uncertainty relation can be interpreted as a bound for the entropy production in terms of current fluctuations \cite{barato2015,gingrich2016,pietzonka2016,marsland2019,horowitz2020}. As shown in Ref. \cite{pietzonka2023}, the statistics of general counting observables bound the entropy production as well. Assuming a specific underlying Markov network, optimization procedures yield tight network specific entropy bounds \cite{teza2020,ehrich2021,skinner2021_1,bisker2023}.

\begin{figure}[bt]
    \begin{subfigure}[t]{0.47\textwidth}
      \centering
        \includegraphics[width=\linewidth]{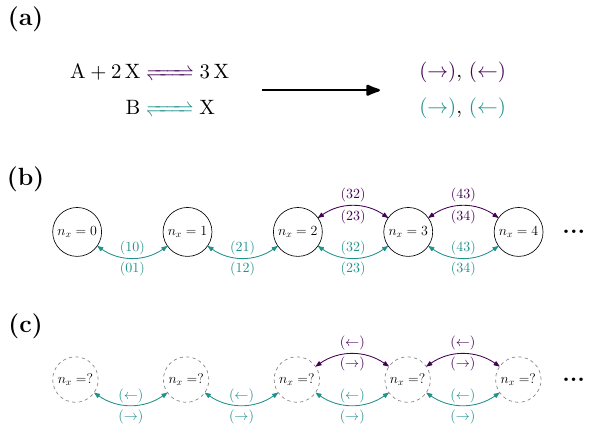}
    \end{subfigure}
    \caption[Schloegel example]{\add{Blurred observation of the Schlögl model. (a) Reaction scheme: $X$ molecules can undergo chemical reactions with $A$ molecules and $B$ molecules provided by two different chemical reservoirs. (b): If an observer resolves the number of $X$ molecules $n_X$, each change in $n_X$ can be represented as transition in an effective Markov network with two different channels. (c): If the observer only registers forward and backward reactions along the A-reaction channel (dark color) or the B-reaction channel (bright color) but cannot resolve the changes in the number of $X$ molecules $n_X$, the corresponding transitions of the effective network are blurred for this observer.}}
\label{Fig:Intro}
\end{figure}

Including waiting times in the effective description, for example via the milestoning coarse-graining scheme \cite{elber2020,hartich2021}, leads to more refined entropy estimators with a broader range of applicability \cite{berezhkovskii2019,martinez2019,hartich2021_2,hartich2021_comment,bisker2022_Reply,skinner2021_2}. In particular, the effective description based on the waiting times between two observable transitions is central for thermodynamic inference as it provides a tight entropy bound \cite{PRX,harunari2022} and additionally contains topological information about the underlying system \cite{li2013,berezhkovskii2020,thorneywork2020,PRX,IJMS}. \add{This inference strategy is based on interpreting the observed transitions as renewal events of an effective semi-Markov description \cite{qian2007,maes2009,PRE} and identifying these renewal events as Markovian events \cite{PRL,PRR}.} 

In realistic scenarios, e.g., the scenarios discussed in Ref. \cite{makarov2023_1}, imperfect measurements can limit the resolution of observations resulting in non-Markovian events. For an example, consider the observation of a chemical reaction network. Instead of observing the precise number of each chemical species, an external observer can potentially only register the type of an observed reaction. Consequently, the individual transitions in the state space are not distinguishable. Stated differently, for this observer, the transitions of the corresponding effective Markov network are blurred. \add{This blurred observation of a chemical reaction network is illustrated in \autoref{Fig:Intro} for the Schlögl model, a paradigmatic example of a chemical reaction network \cite{schloegl1972,matheson1975,vellela2008,remlein2024}.} 

\add{A biophysical example with blurred transitions is an observation of the KaiC-cycle \cite{vanzon2007,johnson2011,paijmans2017,nguyen2018} in which the observer only registers activation and deactivation of KaiC molecules but cannot distinguish between the different molecular states. An example from active matter is an observation of the motion of a run-and-tumble particle subject to thermal noise \cite{schnitzer1993,tailleur2008,pietzonka_2017} in which the orientations of the director are not accessible and therefore blurred. Another example is an observation of a quantum-dot system \cite{sanchez2010,bulnes2011,esposito2012,bulnes2015,borrelli2015} with multiple indistinguishable reservoirs as introduced in Ref. \cite{harunari2024}. Since the observer cannot distinguish which reservoir populates or depletes the system, the corresponding transitions are blurred. Note that from a conceptual point of view, the multifilar events introduced in Ref. \cite{harunari2024} correspond to blurred transitions.}    

This work aims at extending the transition-based description of partially accessible Markov networks to this class of observations. \add{Based on an information-theoretic bound, we derive an operationally accessible entropy estimator which uses blurred transition statistics and the distributions of waiting times between two consecutive blurred transitions.} Crucially, the derived estimator is not the generalization of the entropy estimator for fully resolved transitions since the Markovian event property breaks down for blurred transitions.

We start with a recapitulation of the basic concepts of transition-based thermodynamic inference for resolved transitions in \autoref{Sec:Setup}. These concepts are extended to blurred transitions in \autoref{Sec:Blurred}. In \autoref{Sec:Bound}, \add{we derive the transition-based entropy estimator starting from the corresponding information-theoretic bound.} Various examples illustrating \add{the quality of this estimator} are presented in \autoref{Sec:Examples}. The final \autoref{Sec:Conclusion} contains a concluding perspective.

\section{Setup}\label{Sec:Setup}

\subsection{Underlying Markovian description}
Our starting point is a Markov network with $N$ states for which transitions between state $i$ and state $j$ happen with a time-independent transition rate $k_{ij}$ along a corresponding edge or, equivalently, link of the network. To ensure thermodynamic consistency, we assume \add{that} $k_{ij} > 0$ implies $k_{ji} > 0$. \add{Since} all transition rates are time-independent, the system reaches a stationary state with stationary probabilities $p_i^s$ in the long-time limit $t\to \infty$. If the detailed balance relation $p^s_i k_{ij} = p^s_j k_{ji}$ is broken for at least one edge, this steady state is a non-equilibrium stationary state (NESS) with stationary entropy production rate
\begin{equation}
    \braket{\sigma} = \sum_{ij} p^s_i k_{ij}\ln\frac{p^s_i k_{ij}}{p^s_j k_{ji}} \geq 0,
    \label{EQ:Setup_MarkovEP}
\end{equation}  
where the summation includes all possible transitions \cite{seifert2012}. From a topological perspective, $\braket{\sigma}$ is caused by currents $j_{\mathcal{C}}$ along closed directed loops without self-crossing, i.e., along the cycles $\mathcal{C}$ of the network. Operationally, $j_{\mathcal{C}}$ corresponds to the mean net number of cycle completions divided by the observation time \cite{hill1989,jiang2004}. The contribution of each cycle $\mathcal{C}$ to the entropy production of the network is quantified by the cycle affinity
\begin{equation}
    \mathcal{A}_{\mathcal{C}} \equiv \ln \prod_{ij\in \mathcal{C}}\frac{k_{ij}}{k_{ji}},
    \label{EQ:Setup_MarkovEP_C}
\end{equation}
where the product includes all forward transition rates of the cycle in the numerator and the corresponding backward transition rates in the denominator. Combining the contributions of all cycles of the network, $\braket{\sigma}$ can be calculated via
\begin{equation}
    \braket{\sigma} = \sum_{\mathcal{C}} j_{\mathcal{C}}\mathcal{A}_{\mathcal{C}}
    \label{EQ:Setup_MarkovEP_2}
\end{equation}
which is equivalent to \autoref{EQ:Setup_MarkovEP}. In unicyclic networks, \autoref{EQ:Setup_MarkovEP_2} reduces to
\begin{equation}
    \braket{\sigma} = j_{\mathcal{C}}\mathcal{A}_{\mathcal{C}}.
    \label{EQ:Setup_MarkovEP_UC1}
\end{equation}
For this topology, the cycle current is identical to the current through each link and therefore given by
\begin{equation}
    j_{\mathcal{C}} = p^s_i k_{ij} - p^s_j k_{ji},
    \label{EQ:Setup_MarkovEP_UC2}
\end{equation}
where $i$ and $j$ can be any pair of adjacent states.

\subsection{Resolved observation}

\begin{figure}[bt]
    \begin{subfigure}[t]{0.47\textwidth}
      \centering
        \includegraphics[width=\linewidth]{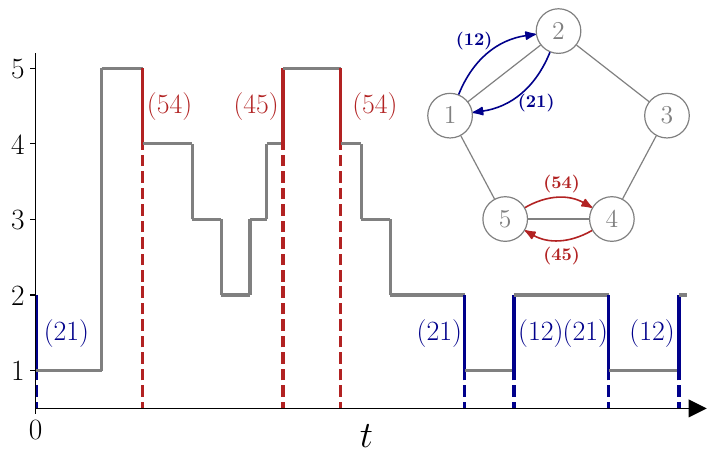}
    \end{subfigure}
    \caption[Resolved trajectory]{Observation of a partially accessible five-state Markov network with four resolved transitions. The resolved observer registers the transitions $(12),(21),$ $(45)$ and $(54)$ and the corresponding in-between waiting times.}
\label{Fig:Resolved}
\end{figure}

We assume that an observer, whom we call ``resolved observer'' for later reference, aims at inferring $\braket{\sigma}$ of the general $N$-state Markov network based on the observation of $2M$ transitions along $M$ different edges. In the course of time, for example during the observation of the five-state Markov network with four resolved transitions shown in \autoref{Fig:Resolved}, the observer registers different transitions and the waiting times between these transitions. This observation results in an effective dynamics for the underlying network that is characterized by waiting time distributions of the form
\begin{equation}
    \psi_{(ij)\to (kl)}(t) \equiv p\left[(kl); T_{(kl)} - T_{(ij)} = t | (ij)\right],
    \label{EQ:Setup_WTD}
\end{equation}
where $T_{(ij)}$ is the time at which transition $(ij)$ is registered. $\psi_{(ij)\to (kl)}(t)$ is the probability density for observing $(kl)$ at time $T_{(kl)} = T_{(ij)} + t$, i.e., after a waiting time $t$, given $(ij)$ was observed at time $T_{(ij)}$. \add{Integrating out the waiting times in \autoref{EQ:Setup_WTD} leads to}
\begin{equation}
    \add{p_{(ij)\to (kl)} = \int_0^{\infty} dt \psi_{(ij)\to (kl)}(t),}
    \label{EQ:Setup_WTD_P}
\end{equation}
\add{which is the probability for observing the resolved transition $(kl)$ after the resolved transition $(ij)$ irrespective of the waiting time in between with normalization $\sum_{(kl)}p_{(ij)\to (kl)} = 1$.}

\begin{figure*}[bt]
    \begin{subfigure}[t]{0.47\textwidth}
      \centering
        \vspace{-6.0cm}
        \includegraphics[width=\linewidth]{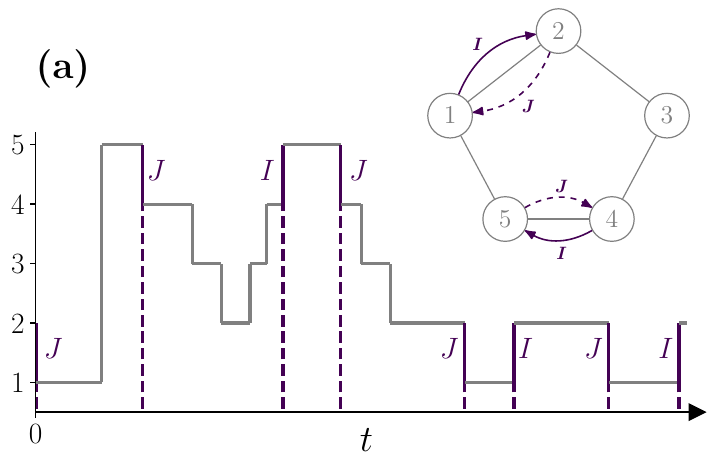}
    \end{subfigure}
    \hfill
    \begin{subfigure}[t]{0.47\textwidth}
      \centering
        \includegraphics[width=\linewidth]{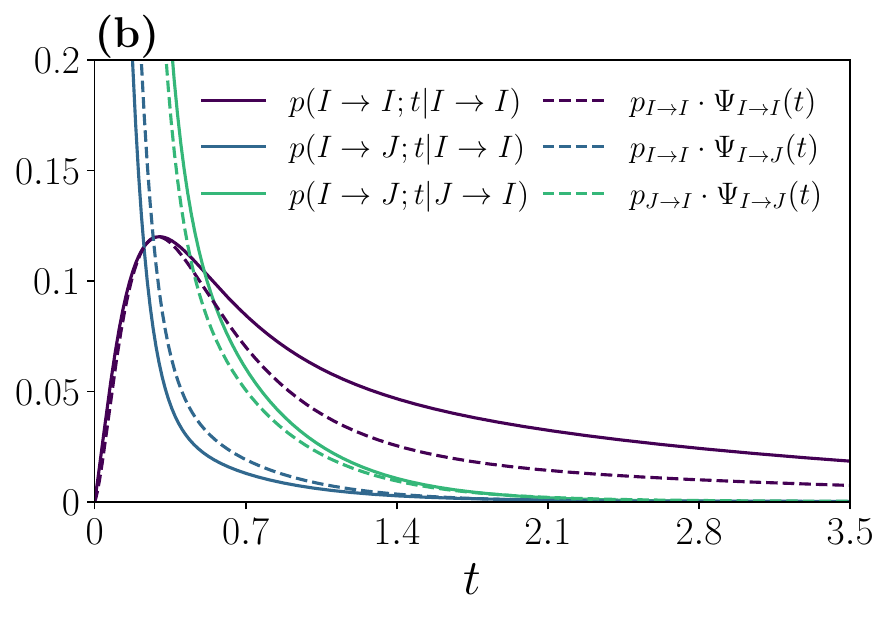}
    \end{subfigure}
    \caption[Blurred trajectory]{Observation of a partially accessible five-state Markov network with two blurred transitions. (a): The blurred observer registers transitions from classes $I$ and $J$ and the corresponding in-between waiting times but cannot resolve the individual transitions within a class. (b): Breakdown of the renewal property, i.e., \autoref{EQ:Indist_WTD_NonMark_2}, for three out of eight possible sequences with three transitions. \add{The deviations between the conditioned probabilities, e.g., $p(I\to J; t|I\to I)$, and the corresponding products, e.g., $p_{I\to I}\cdot\Psi_{I\to J}(t)$, indicate that the observed transitions are not Markovian.} The probabilities and waiting time distributions are calculated with an absorbing master equation for transition rates $k_{12} = 0.4, k_{21} = 8.9, k_{23} = 4.2, k_{32} = 1.1, k_{34} = 1.8, k_{43} = 0.01, k_{45} = 9.4, k_{54} = 0.1,$ $k_{51} = 8.8$ and $k_{15} = 0.6$.}
\label{Fig:Blurred}
\end{figure*}

In general, the emerging effective description of the network is non-Markovian because the described partial observation is not sufficient for determining the state of the underlying network. As a consequence, the corresponding waiting time distributions are typically non-exponential. However, directly after a registered transition, the state of the system is uniquely determined by this transition. For example, in \autoref{Fig:Resolved}, observing $(12)$ implies that the state of the system directly after the transition is two. Therefore, the effective description remains Markovian at the corresponding time instances. 
From a conceptual point of view, these events can be interpreted as the renewal events of a semi-Markov process which describes the effective dynamics in the space of observable transitions \cite{PRX}. \add{Each} observed transition updates the semi-Markov state of the system by renewing the memory of the non-Markovian process with a Markovian event. \add{Thus, in the space of observable transitions, the process is a first-order semi-Markov process.} 

\add{The} updating of the semi-Markov state slices an observed trajectory into snippets starting and ending with an observed transition, i.e., a Markovian event. From an operational point of view, e.g., for the resolved observer, this slicing implies that waiting time distributions for transition sequences with more than two transitions factorize. For example, for the sequence $(ij)\to (kl)\to (mn)$ with in-between waiting times $t_1$ and $t_2$, we have
\begin{equation}
    \psi_{(ij)\to (kl)\to (mn)}(t_1,t_2) = \psi_{(ij)\to (kl)}(t_1)\psi_{(kl)\to (mn)}(t_2).
     \label{EQ:Setup_WTD_Markov}
\end{equation}
\add{Since the memory is renewed at the start and at the end of each snippet, the path weight of the trajectory splits into the path weights of the corresponding subsequent trajectory snippets \cite{PRX,PRL}.} On the level of these trajectory snippets, a fluctuation theorem holds true \add{for their path weights} which relates a single snippet to the completion of paths, especially of cycles \cite{PRX,PRL}. Based on this fluctuation theorem, the entropy estimator
\begin{equation}
    \braket{\hat{\sigma}} \equiv \sum_{(ij), (kl)}\int_0^{\infty}dt \pi_{ij}\psi_{(ij)\to (kl)}(t)\ln\frac{\psi_{(ij)\to (kl)}(t)}{\psi_{(\widetilde{kl})\to (\widetilde{ij})}(t)}
    \label{EQ:Setup_Link_Est}
\end{equation}
for the full entropy production $\braket{\sigma}$ of a partially accessible Markov network can be derived \add{\cite{PRX,harunari2022}}. In \autoref{EQ:Setup_Link_Est}, the summation includes all observed transitions, $(\widetilde{ij}) = (ji)$ is the time-reversed transition of $(ij)$ and $\pi_{ij} = p^s_i k_{ij}$ is the rate for observing transition $(ij)$. This entropy estimator always provides a lower bound on the full entropy production, i.e., $\braket{\hat{\sigma}} \leq \braket{\sigma}$. Especially for unicyclic networks, the observation of transitions $(ij)$ and $(ji)$ along a single link recovers the full entropy production, i.e., $\braket{\hat{\sigma}} = \braket{\sigma} = j_{\mathcal{C}}\mathcal{A}_{\mathcal{C}}$,
because
\begin{equation}
    \int_0^{\infty}dt (\pi_{ij}\psi_{(ij)\to (ij)}(t)-\pi_{ji}\psi_{(ji)\to (ji)}(t))= j_{\mathcal{C}}
    \label{EQ:Uni_j}
\end{equation}
and
\begin{equation}
    \ln\frac{\psi_{(ij)\to (ij)}(t)}{\psi_{(ji)\to (ji)}(t)} = \mathcal{A}_{\mathcal{C}}
    \label{EQ:Uni_a}
\end{equation}
hold true \cite{PRX,harunari2022}. 

\section{From resolved to blurred observations}\label{Sec:Blurred}
We now assume that a second observer with finite resolution, whom we call ``blurred observer'' for later reference, also observes the $2M$ transitions of the underlying network. However, due to the finite resolution of his observation, this observer cannot distinguish between the $2M$ resolved transitions and instead observers $C$ blurred transitions with $2\leq C\leq M$. Stated differently, for the blurred observer, the resolved transitions are grouped into $C$ different effective transition classes $G,H,I,...$ with each transition belonging to one transition class. A concrete example in which the four observable transitions of the network in \autoref{Fig:Resolved} are blurred into two transition classes is shown in \autoref{Fig:Blurred} (a). For these transitions, the blurred observer registers only the corresponding transition class but cannot distinguish between individual transitions within each class. The conditioned probability for observing a specific transition $(ij)$ of transition class $I$ is given by
\begin{equation}
    p(ij | I) = \frac{p^s_i k_{ij}}{\sum_{(kl)\in I}p^s_k k_{kl}} = \frac{\pi_{ij}}{\pi_I},
    \label{EQ:Indist_P_Specific}
\end{equation}
where $\pi_I$ is the rate for observing that a transition within $I$ happens and the summation includes all transitions within transition class $I$. Waiting time distributions for blurred transitions can be interpreted as an average over all waiting time distributions for resolved transitions belonging to the corresponding transition classes. Thus, these waiting time distribution can be calculated via
\begin{equation}
    \Psi_{I\to J}(t) = \sum_{\mathclap{\substack{(ij)\in I\\(kl)\in J}}} p(ij | I)\psi_{(ij)\to (kl)}(t)
    \label{EQ:Indist_WTD_1}
\end{equation}
using the corresponding weight from \autoref{EQ:Indist_P_Specific}. The interpretation of $\Psi_{I\to J}(t)$ is similar to the interpretation of the waiting time distributions for resolved transitions defined in \autoref{EQ:Setup_WTD_Markov}. \add{Note that for the blurred observer, $p(ij | I)$ is not accessible because the individual transitions in a class are not distinguishable. Instead of using \autoref{EQ:Indist_WTD_1}, this observer determines $\Psi_{I\to J}(t)$ directly from recorded histogram data for the corresponding transitions and waiting times.} 

Although the waiting time distributions for blurred transitions can be interpreted analogously to the waiting time distributions for resolved transitions, their meaning on the level of the underlying Markov network is significantly different. Directly after a transition from class $I$, the state of the underlying system can be the final state of any transition within $I$. For example, for the observation scenario shown in \autoref{Fig:Blurred} (a), right after a registered transition from class $I$, the state of the system is either two or five. In general, after a blurred transition, the state of the system is not fully determined, i.e., the state of the system is determined up to $p(ij|I)$. Thus, the effective dynamics on the level of blurred observations is non-Markovian at any point in time.

From the perspective of the blurred observer, this breakdown of the renewal property leads to non-factorizing waiting time distributions, e.g., for the sequence $I\to J\to K$ with waiting times $t_1$ and $t_2$, 
\begin{equation}
    \Psi_{I\to J\to K}(t_1,t_2)\neq \Psi_{I\to J}(t_1)\Psi_{J\to K}(t_2).
    \label{EQ:Indist_WTD_NonMark}
\end{equation}
To illustrate \autoref{EQ:Indist_WTD_NonMark} with the two-transition waiting time distributions of the example in \autoref{Fig:Blurred} (a), we integrate both sides over $t_1$ leading to
\begin{equation}
    p(J\to K;t|I\to J)\neq p_{I\to J}\Psi_{J\to K}(t_2),
    \label{EQ:Indist_WTD_NonMark_2}
\end{equation}
where
\begin{equation}
    p_{I\to J} = \int_0^{\infty} dt \Psi_{I\to J}(t)
    \label{EQ:Indist_WTD_P}
\end{equation}
is the probability to observe a transition from class $J$ after a transition from class $I$ irrespective of the in-between waiting time. For the three different sequences with three transitions shown in \autoref{Fig:Blurred} (b), the renewal property breaks down.

The mapping from resolved transitions to blurred transitions is a time-independent, unique and many-to-one coarse-graining scheme for an already effective description of a partially accessible Markov network by the resolved observer. In the space of observable transitions, this coarse-graining scheme corresponds to lumping effective transition states into effective compound transition states. Crucially, although this strategy is similar to conventional coarse-graining by state lumping \cite{rahav2007,esposito2012,bo2017,seiferth2020}, the behavior of the resulting coarse-grained states under time-reversal is different. If we first apply the time-reversal operation for transitions and blur them into classes afterwards \cite{hartich2021_comment,bisker2022_Reply,PRX}, \add{the blurred transitions of a specific transition class can be even, odd or neither one under time-reversal. We adopt this notion also for the corresponding transition classes, i.e., the lumped states in the space of observable transitions. A transition class $G$ is even under time-reversal if the time-reversed transition of each transition in $G$ is also included in $G$. A transition class $I$ is odd under time-reversal if the time-reversed transition of each transition in $I$ is included in the time-reversed class $\widetilde{I}$. For all other scenarios, the corresponding transition class is neither even nor odd. Note that this notion emphasizes the difference between lumped transition states and conventional lumped states. The latter are always even under time-reversal.} 

As an example, consider again the system shown in \autoref{Fig:Blurred} a. The transition classes $I = \{(12), (45)\}$ and $J = \{(21), (54)\}$ are odd transition classes, i.e., $\widetilde{I} = J$ and $\widetilde{J} = I$ \add{because the time-reversed counterparts of the transitions in $I$ are blurred into $J$}. In contrast, blurring the resolved transitions into $G = \{(12), (21)\}$ and $H = \{(45), (54)\}$ would result in even transition classes, i.e., $\widetilde{G} = G$ and $\widetilde{H} = H$ \add{because the time-reversed counterparts of the transitions in $G$ and $H$ would also be blurred into $G$ and $H$, respectively.} If the transitions were blurred into $K = \{(12), (21), (45)\}$ and $L = \{(54)\}$, the time-reversed transition classes $\widetilde{K} = \{(21), (12), (54)\}$ and $\widetilde{L} = \{(45)\}$ would not be observable \add{and these transition classes would be neither even nor odd under time-reversal. In the following, we assume that for each transition class the corresponding time-reversed one is observable, i.e., we consider only even or odd transition classes.}

%\begin{figure*}[bt]
%    \centering
%    \begin{subfigure}[t]{0.95\textwidth}
%      \centering
%        \includegraphics[width=\linewidth]{Fig3_Cond.pdf}
%    \end{subfigure}
%    \caption[Illustration conditions]{Illustration of the sufficient conditions for applying the entropy bound from \autoref{EQ:Indist_Est_4} to unicyclic networks (a) and multicyclic networks (b). (a) A transition preserving-observation of transition classes $I = \{(12), (45)\}$ and $J = \{(21), (54)\}$ (left) and a non transition-preserving observation of transition classes $G = \{(12), (21), (23), (32)\}$ and $H = \{(45), (54)\}$ (right) of a unicyclic five-state Markov network. (b) A transition-preserving observation of transition classes $G = \{(12), (21)\}$, $H = \{(23), (32)\}$, $I = \{(34), (43)\}$, $J = \{(41), (14)\}$ and $K = \{(13), (31)\}$ including all transitions of the network (left), a transition-preserving observation of transition classes $G = \{(12), (21)\}$, $H = \{(13), (31)\}$, $I = \{(34), (43)\}$ including not all transitions of the network (middle) and a non transition-preserving observation of transition classes $G = \{(12), (21), (23), (32)\}$ and $H = \{(34), (43)\}$ including not all transitions of the network (right) of a multicyclic four-state Markov network.}
%\label{Fig:Conditions}
%\end{figure*}
\null\newpage
\section{Estimation of entropy production}\label{Sec:Bound}
Extending the analogy between waiting time distributions of resolved transitions and blurred transitions to the entropy estimator $\braket{\hat{\sigma}}$ leads to a generalization of \autoref{EQ:Setup_Link_Est} for blurred transitions given by
\begin{equation}
    \braket{\hat{\sigma}_{\mathrm{BT}}} \equiv \sum_{I, J}\int_0^{\infty}dt \pi_{I}\Psi_{I\to J}(t)\ln\frac{\Psi_{I\to J}(t)}{\Psi_{\widetilde{J}\to \widetilde{I}}(t)}
    \label{EQ:Indist_Est}
\end{equation}
where the summation includes all transition classes. Crucially, \add{$\braket{\hat{\sigma}_{\mathrm{BT}}}$} is not a bound for $\braket{\sigma}$, i.e., $\add{\braket{\hat{\sigma}_{\mathrm{BT}}}}\nleq \braket{\sigma}$ \add{in general}, because the entropy estimator for trajectory snippets is not applicable \add{if no renewal events are observed. More specifically, the start and the end of the snippets are then not Markovian events, which breaks the central condition for applying the fluctuation theorem used for deriving $\braket{\hat{\sigma}}$ for resolved transitions \cite{PRX,PRL}.} To derive a waiting time based bound for blurred transitions, we start from \autoref{EQ:Indist_Est} \add{and insert the definition of the waiting time distributions from \autoref{EQ:Indist_WTD_1}. By rewriting the conditioned probabilities using the definition in \autoref{EQ:Indist_P_Specific}, we can separate terms in \autoref{EQ:Indist_Est} and apply the log-sum inequality from information theory \cite{cover2006} which leads to}  
\begin{widetext}
\begin{equation}
    \braket{\hat{\sigma}_{\mathrm{BT}}} + \sum_{I, J}\pi_I p_{I\to J}\log\frac{\pi_I}{\pi_{\widetilde{J}}} \leq \sum_{I, J}\sum_{\mathclap{\substack{(ij)\in I\\(kl)\in J}}}\;\int_0^{\infty}dt \pi_{ij}\psi_{(ij)\to (kl)}(t)\ln\frac{\pi_{ij}\psi_{(ij)\to (kl)}(t)}{\pi_{\widetilde{kl}}\psi_{(\widetilde{kl})\to (\widetilde{ij})}(t)}.\label{EQ:Indist_Est_Add_1}
\end{equation}    
\end{widetext}
\add{For identifying $\pi_I p_{I\to J}$ in the separated term, we have have first carried out the integration over all waiting times in this term and have then used \autoref{EQ:Indist_WTD_1} and \autoref{EQ:Indist_WTD_P}. Note that since we only consider transition classes which are either even or odd under time-reversal, the summation on the right-hand side of the inequality includes the time-reversed partner of each resolved transition. Therefore, we can rewrite \autoref{EQ:Indist_Est_Add_1} as}
\begin{widetext}
\begin{eqnarray}
    \braket{\hat{\sigma}_{\mathrm{BT}}} + \sum_{I, J}\pi_I p_{I\to J}\log\frac{\pi_I}{\pi_{\widetilde{J}}} &\leq & \sum_{I, J}\sum_{\mathclap{\substack{(ij)\in I\\\leq\\(kl)\in J}}}\;\int_0^{\infty}dt (\pi_{ij}\psi_{(ij)\to (kl)}(t) - \pi_{\widetilde{kl}}\psi_{(\widetilde{kl})\to (\widetilde{ij})}(t))\ln\frac{\pi_{ij}\psi_{(ij)\to (kl)}(t)}{\pi_{\widetilde{kl}}\psi_{(\widetilde{kl})\to (\widetilde{ij})}(t)}\\\label{EQ:Indist_Est_Add_2} &\leq & \sum_{(ij)\leq (kl)}\int_0^{\infty}dt (\pi_{ij}\psi_{(ij)\to (kl)}(t) - \pi_{\widetilde{kl}}\psi_{(\widetilde{kl})\to (\widetilde{ij})}(t))\ln\frac{\pi_{ij}\psi_{(ij)\to (kl)}(t)}{\pi_{\widetilde{kl}}\psi_{(\widetilde{kl})\to (\widetilde{ij})}(t)}\\\label{EQ:Indist_Est_Add_Int} &=& \sum_{(ij), (kl)}\int_0^{\infty}dt \pi_{ij}\psi_{(ij)\to (kl)}(t)\ln\frac{\psi_{(ij)\to (kl)}(t)}{\psi_{(\widetilde{kl})\to (\widetilde{ij})}(t)} + \sum_{\mathclap{(ij),(kl)}} \pi_{ij}p_{(ij)\to (kl)}\ln\frac{\pi_{ij}}{\pi_{\widetilde{kl}}}\label{EQ:Indist_Est_Add_4}.
\end{eqnarray}    
\end{widetext}
\add{In the first line, the summation index $(ij)\in I\leq (kl)\in J$ means that each pair of resolved transitions is counted once and that the pairs for which the same transition is observed two times, i.e., $(ij)\to (ij)$, are included. For deducing the inequality in the second line, we have to distinguish two cases. First, the summation in the first line may already include all pairs of resolved transitions in which case the second line is an equality. Second, a pair of resolved transitions may not be included in this summation if $\widetilde{J} = I$ and $\widetilde{I} = J$ holds for the transition classes of these transitions because the corresponding term in \autoref{EQ:Indist_Est} then vanishes. However, the term for this pair of resolved transitions can still be added to the right-hand side of the first line since this term is positive due to $(a-b)\ln a/b \geq 0$ for $a,b \geq 0$. In both cases, the summation in the second line then includes all pairs of resolved transitions. For the third line, we have first rearranged the summation and then separated the contributions of the waiting time distributions from those of the probabilities. For the latter, we have also carried out the integration over the waiting times.} 

\add{After identifying the waiting time contributions as $\braket{\hat{\sigma}}$ based on the definition in \autoref{EQ:Setup_Link_Est} and introducing the abbreviations}
\begin{equation}
	\braket{\hat{\sigma}_{\mathrm{TC}}} = \sum_{I, J}\pi_I p_{I\to J}\log\frac{\pi_I}{\pi_{\widetilde{J}}}
\end{equation}
\add{and}
\begin{equation}
	\braket{\hat{\sigma}_{\mathrm{RT}}} = \sum_{\mathclap{(ij),(kl)}} \pi_{ij}p_{(ij)\to (kl)}\ln\frac{\pi_{ij}}{\pi_{\widetilde{kl}}},
\end{equation}
\add{we obtain from \autoref{EQ:Indist_Est_Add_4} the inequality}
\begin{equation}
    \braket{\hat{\sigma}_{\mathrm{BT}}} + \braket{\hat{\sigma}_{\mathrm{TC}}} \leq \braket{\hat{\sigma}} + \braket{\hat{\sigma}_{\mathrm{RT}}}.
    \label{EQ:Indist_Add_5}
\end{equation}
\add{Note that from a conceptual point of view, the difference $\braket{\hat{\sigma}_{\mathrm{RT}}} - \braket{\hat{\sigma}_{\mathrm{TC}}}$ is related to the entropy of the randomness associated with the different resolved transitions contributing to a blurred transition. As shown in Refs. \cite{esposito2012,bo2017,bo2014}, similar terms also emerge for conventional state-lumping.}

\add{Since we aim at deriving a bound for the full entropy production $\braket{\sigma}$ based on \autoref{EQ:Indist_Add_5}, we have to bound $\braket{\hat{\sigma}_{\mathrm{RT}}}$ because $\braket{\hat{\sigma}}$ is always smaller then $\braket{\sigma}$.} If the underlying Markov network is unicyclic, we plug in \autoref{EQ:Uni_j} and note that
\begin{equation}
    \braket{\hat{\sigma}_{\mathrm{RT}}} = j_{\mathcal{C}}\sum_{(ij), (kl)}\ln\frac{\pi_{ij}}{\pi_{\widetilde{kl}}} \leq j_{\mathcal{C}}\mathcal{A}_{\mathcal{C}} = \braket{\sigma}
    \label{EQ:Indist_Est_5}
\end{equation}
holds true because we can always complete this summation to \autoref{EQ:Setup_MarkovEP_C} by adding $(\pi_{ij} - \pi_{ji})\ln\pi_{ij}/\pi_{ji} = j_{\mathcal{C}}\ln\pi_{ij}/\pi_{ji} \geq 0$ and canceling the steady state probabilities \cite{PRE}. \add{If the underlying network is multicyclic, we reorder the summation in $\braket{\hat{\sigma}_{\mathrm{RT}}}$ which leads to}
\begin{equation}
	\braket{\hat{\sigma}_{\mathrm{RT}}} = \sum_{(ij), (kl)}(\pi_{ij}p_{(ij)\to (kl)} - \pi_{\widetilde{kl}}p_{(\widetilde{kl})\to (\widetilde{ij})})\ln\pi_{ij}.
	\label{EQ:Indist_Add_6}
\end{equation}
\add{Using the normalization of $p_{(ij)\to (kl)}$ and exploiting the stationarity of the NESS, we can rewrite \autoref{EQ:Indist_Add_6} as}
\begin{equation}
	\braket{\hat{\sigma}_{\mathrm{RT}}} = \sum_{(ij)}(\pi_{ij} - \pi_{\widetilde{ij}})\ln\pi_{ij} \leq \braket{\sigma}
	\label{EQ:Indist_Add_7}
\end{equation}
\add{where we have bounded $\braket{\hat{\sigma}_{\mathrm{RT}}}$ by comparing this expression to the definition of $\braket{\sigma}$ in \autoref{EQ:Setup_MarkovEP} and noting that we can always complete the summation by adding $(\pi_{ij} - \pi_{ji})\ln\pi_{ij}/\pi_{ji} \geq 0$. \autoref{EQ:Indist_Add_7} is saturated if all transitions are observed or if $(\pi_{ij} - \pi_{ji})\ln\pi_{ij}/\pi_{ji} = 0$ holds true for all unobserved transitions.} 

\add{With $\braket{\hat{\sigma}_{\mathrm{RT}}} \leq \braket{\sigma}$ for unicyclic and multicyclic networks, \autoref{EQ:Indist_Add_5} reduces to our main result}
\begin{equation}
    \braket{\hat{\sigma}_{\mathrm{BT}}} + \braket{\hat{\sigma}_{\mathrm{TC}}} \leq 2\braket{\sigma},
    \label{EQ:Indist_Est_4}
\end{equation}
\add{which is a bound for the full entropy production of partially accessible Markov networks based on the observation of blurred transitions. This bound can be used as operationally accessible entropy estimator because all quantities entering $\braket{\hat{\sigma}_{\mathrm{BT}}}$ and $\braket{\hat{\sigma}_{\mathrm{TC}}}$ are accessible in a blurred observation. Note that if each transition class includes only one transition, i.e., no transitions are blurred, both sides of \autoref{EQ:Indist_Add_5} are equal, i.e., if $\braket{\hat{\sigma}_{\mathrm{BT}}} = \braket{\hat{\sigma}}$ and $\braket{\hat{\sigma}_{\mathrm{TC}}} = \braket{\hat{\sigma}_{\mathrm{RT}}}$. If the observation additionally contains all transitions of the network, the bound in \autoref{EQ:Indist_Est_4} is saturated because then $\braket{\hat{\sigma}} = \braket{\sigma}$ and $\braket{\hat{\sigma}_{\mathrm{RT}}} = \braket{\sigma}$ hold true.} 

\section{Illustrations}\label{Sec:Examples}
\begin{figure*}[bt]
    \begin{subfigure}[t]{0.325\textwidth}
      \centering
        \includegraphics[width=\linewidth]{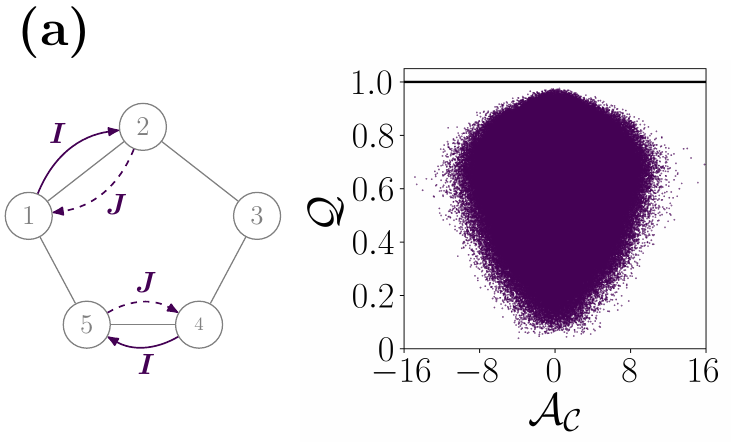}
    \end{subfigure}
    \hfill
    \begin{subfigure}[t]{0.325\textwidth}
      \centering
      \vspace{-3.56cm}
        \includegraphics[width=\linewidth]{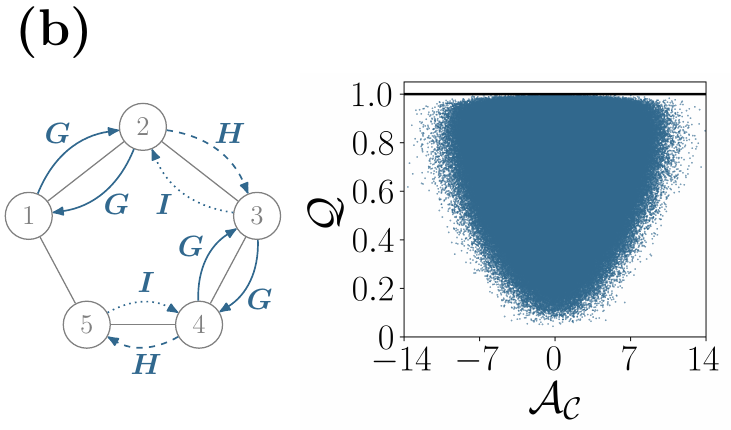}
    \end{subfigure}
    \hfill
    \begin{subfigure}[t]{0.325\textwidth}
      \centering
      \vspace{-3.56cm}
        \includegraphics[width=\linewidth]{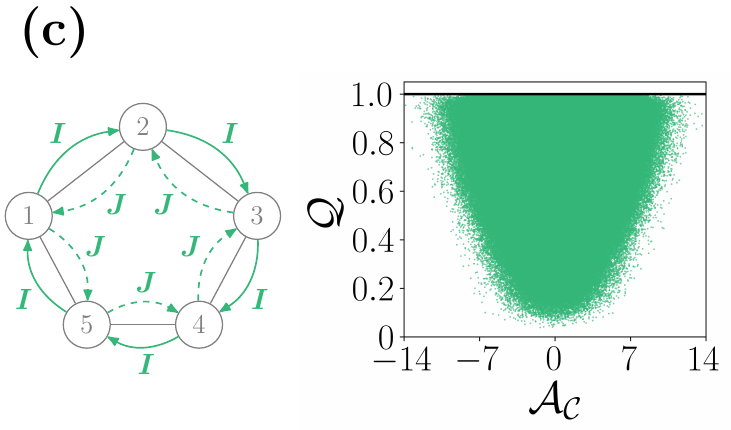}
    \end{subfigure}
    \caption[Quality unicycle]{Quality of the entropy bound in \autoref{EQ:Indist_Est_4} for a unicyclic five-state Markov network as a function of the cycle affinity for three different observation scenarios. (a): Observation of the odd transition classes $I = \{(12), (45)\}$ and $J = \{(21), (54)\}$. The mean quality factor is given by $Q\simeq 0.66$. (b): Observation of the even transition class $G = \{(12), (21), (34), (43)\}$ and of the odd transition classes $H = \{(23), (45)\}$ and $I = \{(32), (54)\}$. The mean quality factor is given by $Q\simeq 0.77$. (c): Observation of the odd transition classes $I$ and $J$ including all forward and backward transitions, respectively. The mean quality factor is given by $Q\simeq 0.8$. Each scenario includes more than 4000000 network realizations with transition rates that are randomly drawn from uniform distributions between $0.1$ and $6$. All waiting time distributions are calculated with an absorbing master equation and \autoref{EQ:Indist_WTD_1}.}
\label{Fig:Q_Uni}
\end{figure*}

We illustrate our results with concrete observation scenarios for four different networks \add{and with the observation scenario for the Schlögl model illustrated in \autoref{Fig:Intro}}. To quantify the quality of the bound in \autoref{EQ:Indist_Est_4}, we introduce the quality factor
\begin{equation}
    \mathcal{Q}\equiv \frac{\add{\braket{\hat{\sigma}_{\mathrm{BT}}}} + \braket{\hat{\sigma}_{\mathrm{TC}}}}{2\braket{\sigma}}\leq 1.
    \label{EQ:Ex_Q}
\end{equation}
\add{We additionally introduce the abbreviation}
\begin{equation}
\Sigma_{\mathrm{BT}}\equiv \frac{\add{\braket{\hat{\sigma}_{\mathrm{BT}}}} + \braket{\hat{\sigma}_{\mathrm{TC}}}}{2}\leq \braket{\sigma} 
\end{equation}
\add{for the entropy estimator derived in \autoref{EQ:Indist_Est_4}.}

\subsection{Unicyclic networks}
We consider three different scenarios for the observation of the unicyclic five-state network from \autoref{Fig:Blurred} \add{shown in \autoref{Fig:Q_Uni}}. For the observation scenario shown in \autoref{Fig:Q_Uni} (a), transitions of two different links are blurred into the odd transition classes $I$ and $J$. For randomly drawn transition rates, the scatter plot shows that $\mathcal{Q} < 1$ for any drawn affinity $\mathcal{A}_{\mathcal{C}}$. \add{In this simulation, the bound is never saturated because only four of ten transitions of the underlying network are observed. Even if \autoref{EQ:Indist_Add_5} is saturated, the inequality in \autoref{EQ:Indist_Est_5} is only saturated for a small set of rates with high symmetry, e.g., for $k_{23} = k_{32}, k_{34} = k_{43}, k_{51} = k_{15}$, which are unlikely to be drawn randomly.} In the modified scenario shown in \autoref{Fig:Q_Uni} (b), transitions of three different links are blurred into the even transition class $G$ and into the odd transition classes $H$ and $I$. The scatter plot shows that $\mathcal{Q} \leq 1$. Compared to the scenario in (a), the quality of the bound increases because only the transitions $(51)$ and $(15)$ are not observed \add{implying that \autoref{EQ:Indist_Est_5} can be saturated for a larger set of rates, e.g., for $k_{51} = k_{15}$, which are more likely to be drawn randomly.} In the observation scenario shown in \autoref{Fig:Q_Uni} (c), all forward transitions are blurred into the odd transition class $I$ and all backward transitions are blurred into the odd transition class $J$. Since this observation includes all transitions of the network on the resolved level, \add{i.e., $\braket{\hat{\sigma}_{\mathrm{RT}}} = \braket{\sigma}$,} the bound can be saturated \add{for all rates that saturate \autoref{EQ:Indist_Add_5}}. 

\subsection{Multicyclic networks}

\begin{figure*}[bt]
    \begin{subfigure}[t]{0.325\textwidth}
      \centering
        \includegraphics[width=\linewidth]{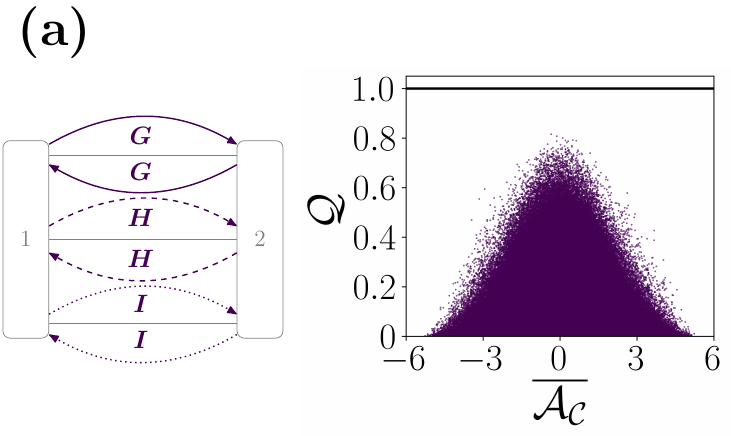}
    \end{subfigure}
    \hfill
    \begin{subfigure}[t]{0.325\textwidth}
      \centering
      \vspace{-3.56cm}
        \includegraphics[width=\linewidth]{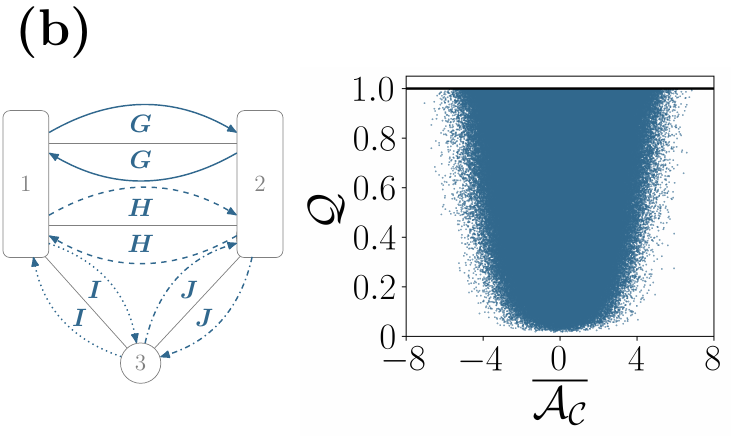}
    \end{subfigure}
    \hfill
    \begin{subfigure}[t]{0.325\textwidth}
      \centering
      \vspace{-3.56cm}
        \includegraphics[width=\linewidth]{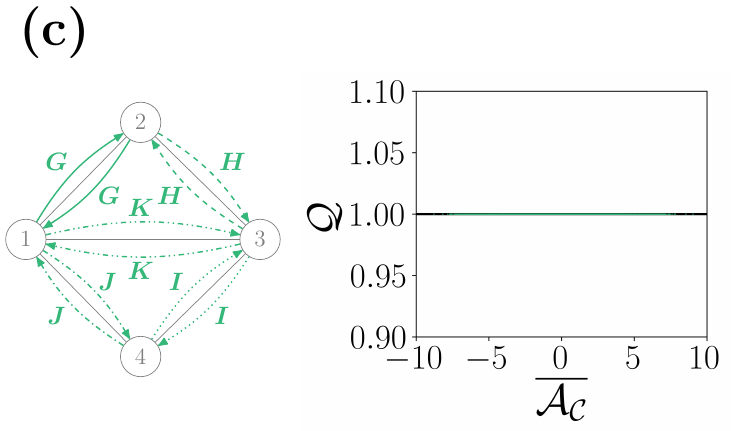}
    \end{subfigure}
    \caption[Quality multicycle]{Quality of the entropy bound in \autoref{EQ:Indist_Est_4} for the observation of three multicyclic Markov network as a function of the average cycle affinity $\overline{\mathcal{A}_{\mathcal{C}}}$. For each link, the forward and backward transitions are blurred into the same transition class. (a): Observation of a three channel network with even transition classes $I, J$ and $K$. The mean quality factor is given by $Q\simeq 0.06$. (b): Observation of a three state network with two channels and even transition classes $H, I, J$ and $K$. The mean quality factor is given by $Q\simeq 0.67$. (c): Observation of a four state network with even transition classes $G, H, I, J$ and $K$. The mean quality factor is given by $Q\simeq 1.0$. Each observation includes more than 1300000 network realizations with transition rates that are randomly drawn from uniform distributions between $0.1$ and $6$. All waiting time distributions are calculated with an absorbing master equation and \autoref{EQ:Indist_WTD_1}.}
\label{Fig:Q_Multi}
\end{figure*}

We consider the observation of three different multicyclic networks shown in \autoref{Fig:Q_Multi} (a)-(c). The networks in (a) and (b) have multiple channels for the transitions $(12)$ and $(21)$. For each network, all transitions are observed on the resolved level and the forward and backward transitions of each link or each channel are blurred into a single even transition class. The corresponding scatter plots show that $\mathcal{Q}$ as a function of the average cycle affinity $\overline{\mathcal{A}_{\mathcal{C}}} = \sum_{\mathcal{C}_i}\mathcal{A}_{\mathcal{C}_i}/(\#\;\mathrm{cycles})$, i.e., the sum of all cycle affinities divided by the number of cycles, is always smaller than one or equal to one.

The mean quality factors for the observation in \autoref{Fig:Q_Multi} (a), $Q\simeq 0.06$, in \autoref{Fig:Q_Multi} (b), $Q\simeq 0.67$, and in \autoref{Fig:Q_Multi} (c), $Q\simeq 1.0$, are significantly different indicating distinct quality regimes of the bound. Since, as previously mentioned, the bound is saturated if no transitions are blurred, i.e., if each registered transition determines the state of the underlying system completely, one possible explanation for this result is the different degree of non-Markovianity of the considered scenarios. In the scenario shown in \autoref{Fig:Q_Multi} (a), each registered blurred transition is non-Markovian because each pair of subsequent transition classes has two possible realizations on the level of resolved transitions. In contrast, in the scenario shown in \autoref{Fig:Q_Multi} (b), registering the blurred transitions $I\to J$ or $J\to I$ corresponds to a Markovian event because, due to the topology of the network, the only possible resolved transition sequences for $I\to J$ and $J\to I$ are $(13)\to (32)$ and $(23)\to (31)$, respectively. Generalizing this argumentation to the scenario in \autoref{Fig:Q_Multi} (c), the bound in \autoref{EQ:Indist_Est_4} is then saturated because for this network topology, each observed blurred transition corresponds to one unique transition registered by the resolved observer. Thus, each observed blurred transition is a Markovian event and the full $\braket{\sigma}$ is recovered in such a scenario where transitions of all links are observed without resolving their directionality. 

\subsection{\add{Schlögl model}}

\add{We consider the blurred observation of the Schlögl model as introduced in \autoref{Fig:Intro}. Following the reaction scheme \cite{schloegl1972,matheson1975,vellela2008}}
\begin{eqnarray}
    \ce{A + 2X}&\ce{<=>[{k_+^{\mathrm{A}}}][{k_{-}^{\mathrm{A}}}] 3X}\\
    \ce{B}&\ce{<=>[{k_+^{\mathrm{B}}}][{k_{-}^{\mathrm{B}}}] X},\label{Eq:Schloegl}
\end{eqnarray}
\add{where $k_{\pm}^{\mathrm{A}}$ and $k_{\pm}^{\mathrm{B}}$ are the forward and backward transition rates of the A-reaction channel and the B-reaction channel, respectively, $X$ molecules contained in an open system with size $\Omega$ can undergo chemical reactions with $A$ and $B$ molecules provided by two external reservoirs at fixed concentrations $c_A$ and $c_B$. By the laws of mass-action kinetics, the transition rates of these reactions are determined by molecular rate constants via}
\begin{eqnarray}
   k_+^{\mathrm{A}}(n_X) &=& \kappa_+^{A}c_A\frac{n_X (n_X-1)}{\Omega}\\
   k_-^{\mathrm{A}}(n_X) &=& \kappa_-^{A}\frac{n_X (n_X-1)(n_X-2)}{\Omega^2}\\
   k_+^{\mathrm{B}}(n_X) &=& \kappa_+^{B}c_B\Omega\\
   k_-^{\mathrm{B}}(n_X) &=& \kappa_-^{B}n_X,\label{Eq:Schloegl_Rates}
\end{eqnarray}
\add{where $n_X$ is the number of $X$ molecules.}

\add{We assume that the system is out of equilibrium due to a difference in chemical potential $\Delta\mu = \mu_B - \mu_A$ between the two chemical reservoirs. For thermodynamic consistency, we require that the local detailed balance condition}
\begin{equation}
    \Delta\mu = \ln\frac{c_A \kappa_+^{A} \kappa_-^{B}}{c_B \kappa_-^{A} \kappa_+^{B}}
    \label{Eq:Schloegl_Aff}
\end{equation}
\add{holds true. By transforming one $A$ molecule into one $B$ molecule via the creation and depletion of one $X$ molecule, the system completes a cycle with affinity $\Delta\mu$. These cycle completions lead to a non-vanishing mean entropy production rate}
\begin{equation}
	\braket{\sigma} = j_{X}\Delta\mu
	\label{Eq:Schloegl_Ent}
\end{equation}
\add{where $j_{X}$ is the mean current of molecules entering and leaving the two reservoirs.}   

\begin{figure}[bt]
    \begin{subfigure}[t]{0.36\textwidth}
      \centering
        \includegraphics[width=\linewidth]{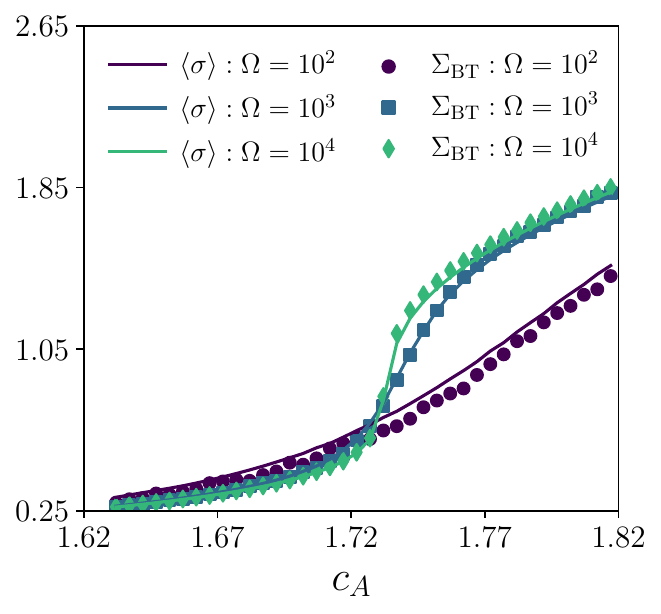}
    \end{subfigure}
    \caption[Schloegel]{\add{Entropy production rate $\braket{\sigma}$ and entropy estimator $\Sigma_{\mathrm{BT}}$ for the blurred observation of the Schlögl model as a function of $c_A$ at different system sizes $\Omega$ with $\Delta\mu = \ln 9$ and $\kappa_\pm^{A/B} = 1$. The stationary distribution of $n_X$ needed for calculating the waiting time distributions via \autoref{EQ:Indist_WTD_1} is estimated numerically based on Gillespie simulation data. The slight deviations between $\braket{\sigma}$ and $\Sigma_{\mathrm{BT}}$ emerge as a consequence of the numerical error of the estimation method for the stationary distribution and as a consequence of the finite sample size. The increase in $\braket{\sigma}$ at $c_A \simeq 1.73$ is caused by the phase transition of the Schlögl model at the critical point \cite{matheson1975,vellela2008}.}}
\label{Fig:Ent_Schloegl}
\end{figure}

\add{Following the observation scenario in \autoref{Fig:Intro}, we assume that an external observer with no access to $n_X$ aims at inferring $\braket{\sigma}$ based on the observation of the different reaction types $(A\rightarrow), (A\leftarrow), (B\rightarrow)$ and $(B\leftarrow)$. As shown in \autoref{Fig:Ent_Schloegl}, the bound in \autoref{EQ:Indist_Est_4} saturates within numerical errors for all considered concentrations $c_A$ and system sizes $\Omega$ which implies that the estimator $\Sigma_{\mathrm{BT}}$ recovers the full entropy production of the system for these parameters.}

\add{This numerical result can be rationalized by comparing the amount of information accessible for the blurred observer to the amount of information contained in $\braket{\sigma}$ in \autoref{Eq:Schloegl_Ent}. Although the effective Markov network for the resolved observation of the Schlögl model illustrated in \autoref{Fig:Intro} (b) is multicyclic, \autoref{Eq:Schloegl_Ent} does not resolve each cycle of the network individually because, due to symmetry, all cycles of the network have the same affinity $\Delta\mu$. Furthermore, $j_{X}$ is independent of resolving individual cycles because this current recovers the average number of created and consumed $A$ and $B$ molecules of the whole network. Conceptually, the same amount of information is included in $\Sigma_{\mathrm{BT}}$. Each blurred transition corresponds to the creation or the consumption of one $A$ or one $B$ molecule recovering the information contained in $j_X$. Additionally, as shown in general and illustrated for the Brusselator model in Ref. \cite{harunari2024}, the probabilities for specific blurred transition sequences, for example $(A\rightarrow)\to (B\leftarrow)$ or $(A\leftarrow)\to (B\rightarrow)$, recover the cycle affinity $\Delta\mu$. Thus, the blurred observation captures the information contained in $\braket{\sigma}$ implying the saturation of the corresponding entropy bound.}              

\section{Conclusion}\label{Sec:Conclusion}
In this paper, we have extended the transition-based description of partially accessible Markov networks to those with blurred transitions by introducing the concept of transition classes. To establish an effective description from the perspective of such a blurred observer, we have introduced waiting time distributions for transitions between transition classes. Based on these waiting time distributions, we have demonstrated that this effective description is in general non-Markovian, i.e., does not include any renewal event. As a consequence, the extant entropy estimator from Ref. \cite{PRX} defined for resolved observed transition is not applicable. As this result implies that the formulation of any direct waiting time based entropy estimator most likely will fail, we have proven a complementary bound based on the log-sum inequality which reduces to an operationally accessible entropy estimator. Furthermore, we have illustrated the operational relevance of this \add{estimator} with various examples. 

Future work could address the following issues. So far, the effective description is formulated for partially accessible Markov networks, which implies that its range of applicability is restricted to discrete systems. Generalizing the concept of transition classes to continuous degrees of freedom is \add{therefore} a possible next step. This generalization could additionally lead to an entropy estimator for the continuous analogue of blurred transitions. Similarly, the description can potentially be generalized to systems which are not in a NESS, for example to systems with time-dependent driving. Our coarse-graining procedure leads to another open question. As state lumping is only one established coarse-graining scheme out of many, the extension of other schemes to the space of observable transitions might be possible as well. Assuming that these extensions lead to analogous bounds with a different range of validity, it might be possible to infer the realized coarse-graining in an observed systems based on the saturation or violation of the respective bounds.\\ 

\section*{Acknowledgments}
\add{We thank Julius Degünther and Benedikt Remlein for fruitful discussions, the latter for providing raw simulation data for the Schlögl model, and Jann van der Meer for insightful discussions and a crucial moment of inspiration.}

%\nocite{*}

%\bibliography{references.bib}
%\bibliographystyle{apsrev4-2}


\end{document}